\documentclass[
    ,final            
  ,numberedheadings 
  ]
  {aipproc}
\usepackage{graphicx,pstricks,subfigure}
\layoutstyle{8x11single}

\usepackage{amsmath,amssymb}

\newcommand{\f}{\frac}
\newcommand{\suml}{\sum\limits}
\newcommand{\m}{\mathbf}
\newcommand{\intl}{\int\limits}

\newcommand{\eql}[1]{Eq.~\eqref{#1}}
\newcommand{\eqls}[2]{Eqs.~\eqref{#1},\,\eqref{#2}}
\newcommand{\bs}{\boldsymbol}

\graphicspath{{figures/}}

\begin{document}

\title{The kinetic regime of the Vicsek model}


\author{A. A. Chepizhko}{address={Department for Theoretical Physics, Odessa National University, Dvoryanskaya 2, 65026 Odessa, Ukraine}
}

\author{V. L. Kulinskii}{address={Department for Theoretical Physics, Odessa National University, Dvoryanskaya 2, 65026 Odessa, Ukraine}}

\begin{abstract}
We consider the dynamics of the system of self propelling particles modeled via the Vicsek algorithm in continuum time limit. It is shown that the alignment process for the velocities can be subdivided into two regimes: ``fast`` kinetic and ``slow`` hydrodynamic ones. In fast kinetic regime the alignment of the particle velocity to the local neighborhood takes place with characteristic relaxation time. So that the bigger regions arise with the velocity alignment. These regions align their velocities thus giving rise to hydrodynamic regime of the dynamics. We propose the mean-field like approach in which we take into account the correlations between density and velocity. The comparison of the theoretical predictions with the numerical simulations is given. The relation between Vicsek model in the zero velocity limit and the Kuramoto model is stated. The mean-field approach accounting for the dynamic change of the neighborhood is proposed. The nature of the discontinuity of the dependence of the order parameter in case of vectorial noise revealed in Gregorie and Chaite, Phys. Rev. Lett., {\bf 92}, 025702 (2004) is discussed and the  explanation of it is proposed.
\end{abstract}

\maketitle


\section*{Introduction}
The equilibrium statistical mechanics and thermodynamics of
Hamiltonian systems are well developed areas of Statistical
Physics. There are also a lot of remarkable results for open
systems, which considered far from equilibrium
\cite{book_haken_synergy,book_prigoginenicolis}. In general one
can consider the systems with some constraints which bounds the
coordinates of the particles since the general formalism
remains unchanged (e.g. Liouville equation etc.). It is well
known that due to instability of trajectories the Hamiltonian
systems reach the thermodynamic equilibrium which can be
characterized several macroscopic parameters of state despite
huge number of microscopic degrees of freedom. In addition
total momentum and angular momentum are conserved. Though for
every specific configuration the formation of stationary local
vortical structures
\cite{spp_eberdmikh_pre2005,spp_bertoz_prl2006,spp_china_pre2006}
may occur due to conservation of the angular momentum.
Obviously the inclusion of potential forces has little grounds
for the systems of intellectual particles (individuals in
flocks, crowds etc.), which differ very much in this respect
from the molecular system for which either all forces have
mainly potential character or dissipative.

In \cite{spp_cva_prl1995} the minimal model, the so called
Vicsek model (VM) of such type was introduced. The dynamic rule
for the alignment of the particle' velocities constructed in
such a way that at high density the kinetic energy of
disordered motion is transformed into the one of ordered motion
so that the total kinetic energy is conserved. Then the system
reaches the final state with nonzeroth total momentum even in
the low (one- and two-) dimensional cases. In such a case the
appearance of the ordered state is predetermined by the dynamic
rule. Note that the VM does not take into account the potential
interparticle forces.

In this paper we consider the kinetic regime for the VM when the particle aligns along the velocity direction of its neighborhood and give the estimation for the critical noise amplitude of order-disorder transition. The structure of paper is as follows. In Section~\ref{sec_1} we derive the continuum time analog of the VM and show that angular velocity of a particle consists of two terms which describe alignment. One of the terms describes the fast kinetic relaxation to the local direction the other one describes the hydrodynamic regime of alignment between the domains where local alignment is settled down. In Section~\ref{sec_2} the process of the relaxation of one-particle velocity to the local value of the neighborhood is considered. The dependence of the rate of the relaxation on the density is obtained by the numerical experiment and the corresponding Fokker-Plank equation is derived. In Section~\ref{sec_3} the influence of the dynamic nature of the environment is discussed. The results obtained are summarized in the conclusion.

\section{Vicsek model in continuum time limit and the two regimes of the dynamic}\label{sec_1}
The Vicsek model of the dynamic of self-propelling particles
\cite{spp_cva_prl1995} can be represented by the relation:
\begin{equation}\label{eq_cva0}
\m{v}_{i}\left(n+1\right)\times\m{u}_{i}\left(n\right)=0,\quad\forall\, i,n\,,
\end{equation}
Here $\mathbf{u}_{i}(n)$
\begin{equation}
\mathbf{u}_{i}=\f{\suml_{j}H(\mathbf{r} _{ij})\,\m{v}_{j}}{
\left\vert \suml_{j}H(\mathbf{r} _{ij})\,\m{v}_{j} \right\vert }
\label{u}
\end{equation}
is the unit vector corresponding to the averaged velocity of
the neighborhood and $H(r_{ij})$ is the averaging kernel with
the characteristic averaging scale $R$. The absolute value of
the velocity of each particle is assumed to be constant, i.e.
\begin{equation}
\mid\m{v}_{i}\left(n+1\right)\mid=\mid\m{v}_{i}\left(n\right)\mid=\it{v}_{i}\,.
\end{equation}
The noise is not included. For the VM, $H$ is proportional to a
Heaviside step function. One can also consider other models for
the averaging kernel. One can say that the dynamics of
individual particle is subjected to reduction of the difference
between the direction of its velocity and that of the average
velocity of the surrounding given by \eql{u}.

Note that at every step the direction of the velocity
$\m{v}(n+1)$ coincides exactly with the direction of
$\m{u}(n)$. Another words, the vector $\m{u}(n)$ remains
unchanged during the velocities updating. This is specific for
discrete formulation but this is not the case in the continuous
time limit since both vectors are rotating during infinitesimal
time interval $\delta t$. In such a limit the angular velocity
$\bs{\omega}_{\m{v}}$ consists of two parts. The first one is
the angular velocity $\bs{\omega}_{\m{u}}$ of the unit vector
$\m{u}$ for the average velocity of the nearest neighbors. The
second one is the relative angular velocity. When the time is
continuous taking into account constraint \eql{eq_cva0} the
equation of motion of such a particle can be written as:
\begin{equation}\label{eq_eom}
\frac{d}{dt}\m{v}_{i}=\bs{\omega}_{\m{v}_i}\times \m{v}_{i}\,\,.
\end{equation}
Here $\bs{\omega}_{\m{v}_i}$ is the ``angular velocity`` of $i$-th particle.

This angular velocity depends on the velocities of neighboring
particles. The self-propelling force and the frictional force
are assumed to balance each other. The hydrodynamic model which
is based on the equations of motion \eqref{eq_cva0} and it
continual analog \eqref{eq_eom} was proposed in
\cite{spp_us_eurphyslet2005,spp_usphysicastacsolut_physica2006}.

Note that in the discrete CVA at every step the direction of
the velocity $\m{v}(n+1)$ coincides exactly with the direction
of $\m{u}(n)$. Another words, the vector $\m{u}(n)$ remains
unchanged during the velocities updating. This is specific for
discrete formulation but this is not the case in the continuous
time limit since both vectors are rotating during infinitesimal
time interval $\delta t$. In such a limit the angular velocity
$\bs{\omega}_{\m{v}}$ consists of two parts. The first one is
the angular velocity $\bs{\omega}_{\m{u}}$ of the unit vector
$\m{u}$ for the average velocity of the nearest neighbors. The
second one is the relative angular velocity.
\begin{equation}\label{om}
  \bs{\omega}_{\m{v}_{i}} = \bs{\omega}_{\m{u}_{i}} +
  \bs{\omega}_{\m{v}\m{u}}\,,
\end{equation}
where
\begin{align}
\bs{\omega}_{\m{u}_{i}} =&\, \m{u}_{i}\times\dot{\m{u}}_i \label{om12}\\
\bs{\omega}_{\m{v}\m{u}} =&\, A\,\m{v}_{i}\times\m{u}_{i}\,.
\label{om13}
\end{align}
The quantity $A>0$ is inverse to characteristic length scale.
The latter is the radius of interaction $R$ and is the
parameter of the model. Indeed, in the limit $R\to \infty$
each particle has the same neighborhood, provided that $N\gg1$,
i.e. $\m{u}_{i}$ does not depend on $i$. Therefore, in such a
limit they all has the same angular velocity, which is given by
the first term of \eql{om}. From \eql{om} and \eqls{om12}{om13}
for 2D case we obtain:
\begin{equation}\label{omi}
\dot{\bs{\omega}}_{\m{v}_{i}} = - A \left( \m{v}_{i}\cdot\m{u}_{i}
\right)\, (\bs{\omega}_{\m{v}_{i}}-\bs{\omega}_{\m{u}_{i}}) -
(\dot{\bs{\omega}}_{\m{u}_{i}}+\bs{\omega}_{\m{u}_{i}}^2)\,.
\end{equation}

\section{One-particle relaxation}\label{sec_2}
Here we show that the equation \eqref{om} in 2D is closely
related to well known Kuramoto model (KM) for the phase
synchronization. Indeed, let the angle $\theta_i$ characterize
the direction of the velocity of $i$-th particle, then
$\bs{\omega}_{\m{v}_{i}}= \dot{\theta}_{i}$ and the equation
\eqref{om} takes the form:
\begin{equation}\label{km}
  \dot{\theta}_{i} =\dot{\psi_{i}} +A\,\sin\left(\,\psi_i - \theta_i \,\right)
\end{equation}
Here $\psi_i$ denotes the angle which determines the direction
vector $\m{u}_i$. It is one of the variant of the short-range
version of the KM \cite{spp_kuramotosakaguchi_japan1987} (see
also \cite{spp_kuramotomodel_rpmreview2005} and reference
therein) of the form:
\begin{equation}\label{kmlocal}
  \dot{\theta}_{i} =\omega_{i} +
  K\,\suml_{\left\langle\,i,j\,\right\rangle }\sin\left(\,\theta_j - \theta_i
  \,\right)\,,
\end{equation}
where the brackets stand for nearest-neighbor oscillators. Thus
we can state that in the zero velocity limit the Vicsek model
with continuum time belongs to the short-range Kuramoto model
class \cite{spp_kuramotomodel_rpmreview2005}. This allow to
conclude that for low velocity the ordering in the Vicsek model
is governed by the same mechanism as the synchronization in the
KM. Since the synchronization transition is of continuum type
one can expect that the continuum character of the transition
take place for low enough velocity in Vicsek model too. This
conclusion is in correspondence with the results of
\cite{spp_vicseknagy_pha2007}.

According to its definition vector $\m{u}_{i}$ changes slower
than the velocity of a particle $\m{v}_i$. From Eq.~\eqref{omi}
it follows that the second term in \eql{om} governs the kinetic
of the alignment process while the first term is of
hydrodynamic character since it determines the behavior on
scales greater than $R$. Therefore \eql{omi} shows that in
continuous time limit the CVA system has the stable state where
the particles align along some direction, which is
characterized by the director $\m{u}_0$. Equations
\eqref{eq_eom}, \eqref{om} can serve as the continuum time
analog for the CVA. Additional confirmation of that is the
behavior of these terms with respect to reverting the time
$t\to -t$. The first term changes its sign and therefore
produces the reversible contribution to the equation of motion,
while the  second term does not change the sign thus
representing irreversible part of the CVA, which  governs
irreversible one-particle kinetics of the alignment.
%
%
To study Eq.~\eqref{omi} analytically we use the approximation
which takes into account that the variable $\m{u}_i$ is the
``collective`` one, thus there is the time interval which we
call ``kinetic`` regime where it changes much slower so that
$\bs{\omega}_{\m{u}_{i}}$ and its derivative can be omitted. In
addition we assume that the value of $A$ does not depend on
time which reflects that the number of ``interacting``
neighbors remains constant. In such an approximation
Eq.~\eqref{omi} reduces to more simple form:
\begin{equation}\label{omi0}
\dot{\bs{\omega}}_{\m{v}_{i}} = -
A \left( \m{v}_{i}\cdot\m{u}_{i}\,. \right)\, \bs{\omega}_{\m{v}_{i}}\,.
\end{equation}
In scalar form for the angle of the alignment $\alpha_i$
between the vectors $\m{v}_i$ and $\m{u}_i$ taking into account
that $\bs{\omega}_{\m{v}_{i}} = \ddot{\alpha}$ and
\[\m{v}_{i}\cdot\m{u}_{i} = \cos{\alpha}\,,\] after integrating
Eq.~\eqref{omi0} we obtain:
\begin{equation}\label{omialpha}
\dot{\alpha} = - A \sin{\alpha}\,.
\end{equation}
where $A$ is some parameter which determines the alignment rate
and obviously depends on the density and the average velocity
of the neighbors. Here we put the following initial condition
$\dot{\alpha} = 0$, which is in accordance with that in
simulations. The solution of \eql{omialpha} is:
\begin{equation}\label{solalpha}
\tan{\f{\alpha}{2}} =\tan{\f{\alpha_0}{2}}\,\,\exp{\left(\,-A\,t\,\right)}\,, \end{equation}
Thus  the one-particle alignment process has relaxation character.

\begin{figure}[hbt!]
\includegraphics[scale=0.15]{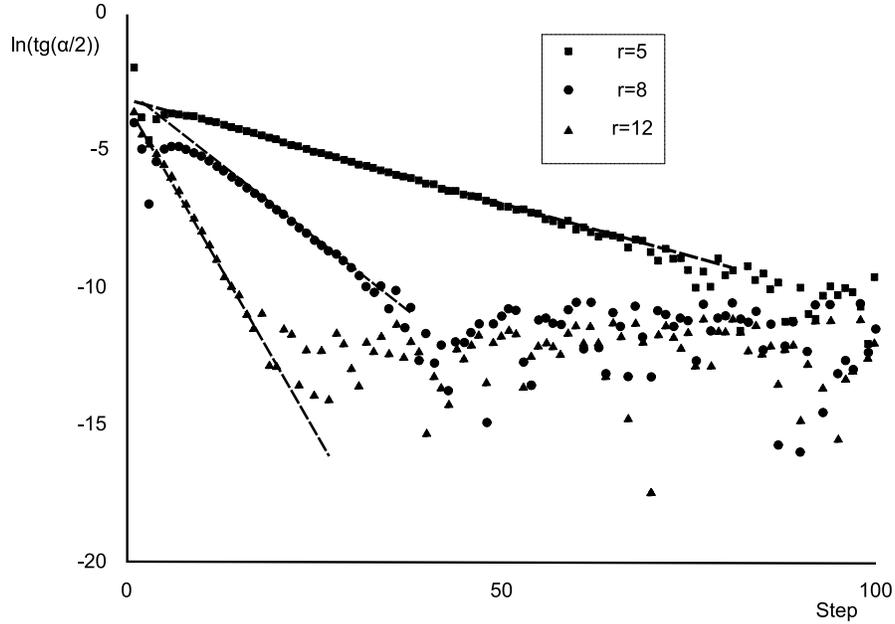}
  \caption{The log plot for the value $\tan{\f{\alpha}{2}}$ as the function of time (number of steps) at different radius of interaction $r$. The results of simulation with $N = 900\,, L = 30\,, v = 0.1$.}\label{angle}
\end{figure}

To compare this result with the simulation data we performed
the simulation with the small initial disalignment of the
directions of the particles in dense system $\rho = N
\,\left(\,\f{1}{L} \,\right)^2 \approx 1 $. The results of
simulation are represented on Fig.~\ref{angle} and demonstrate
the existence of such an interval, where the dependence given
by Eq.~\eqref{omi0} takes place.

\begin{figure}
  \includegraphics[scale=0.75]{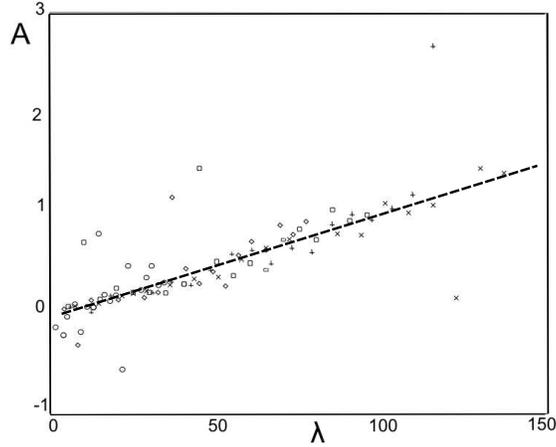}\\
  \caption{The dependence of the relaxation rate on the number of the neighbors $\lambda  = \rho r^2$.}\label{fig_koef_vs_dens}
\end{figure}

The kinetic regime of the system subjected to the stochastic
increment of the direction
\cite{spp_cva_prl1995,spp_cva_pre1996} can be described by
stochastic modification of \eql{omialpha}:
\begin{equation}\label{omialpha1}
\dot{\alpha} = - A \sin{\alpha}+L(t)\,,
\end{equation}
where $L(t)$ is the standard white noise term. Then
\eql{omialpha1} is equivalent to Fokker-Plank equation for the
density distribution function $f_{\m{v}}(\alpha,t)$
\cite{book_vankampen}:
\begin{equation}\label{kinetic_eq_fv}
  \frac{\partial\,f_{\m{v}}}{\partial\, t}\,=\,
  \frac{\partial\, }{\partial\, \alpha }\left(\,A\,\sin \alpha \,f_{\m{v}}
  \,\right)+
  D \,
  \frac{\partial^2\, f_{\m{v}}}{\partial\, \alpha^2}\,,
\end{equation}
where $D$ is the diffusion coefficient and we use the approximation:
\begin{equation}\label{A_model}
A = \lambda \,u
\end{equation}
for the alignment rate for the dimension reasons, where
$\lambda$ is the local density. Such density dependence of the
relaxation rate is supported by the simulations (see
Fig.~\ref{fig_koef_vs_dens}).

The case $D=0$ corresponds to the deterministic case, with the solution:
\begin{equation}\label{deter}
  f^{(0)}_{\m{v}}(\alpha ,t)= \f{f\left(\,e^{\lambda \,u\,t}\tan\f{\alpha }{2} \,\right)}{\sin \alpha
  }\,,
\end{equation}
which demonstrates the alignment process in accordance with
Eqs.~\eqref{omialpha}, \eqref{solalpha}. The stationary
solution of Eq.~\eqref{kinetic_eq_fv} is:
\begin{equation}\label{fpstac}
  f^{(st)}_{\m{v}}(\alpha) = C(D)\,e^{\lambda \,\f{u}{D}\,\cos \alpha }\,.
\end{equation}
Note that the distribution \eqref{fpstac} was used in
\cite{spp_cvalattic_pre1995} for the lattice model as the
analog of the Boltzman distribution. Thus the consideration
presented above can serve as the ground for such
representation.

\section{The account of dynamical environment}\label{sec_3}
In the zero velocity limit the Vicsek model can be considered
in terms of the lattice model as the systems of interacting
spins with the interaction favoring the alignment and the
emergence of the long range order. Yet there is the major
difference between the CVA and the equilibrium models. This is
the coupling between the density and the velocity fields. Due
to such coupling in the static ($v = 0$) case the the
equilibrium systems does not order for densities below some
threshold value close to 1 (which corresponds to the
percolation threshold of randomly distributed spheres) while in
the SSP ordering is found for all velocities
\cite{spp_vicseknagy_pha2007}. The instant change of the
environment in the neighborhood of the particle can be
considered as the noise factor correlated with the local value
of the order parameter - the average velocity of the neighbors:

\begin{equation}\label{localorderparameternorm}
   \m{u}_{i}=\f{1}{N_{i}}\sum _{ \left\langle\, i,j \,\right\rangle }\m{v}_{i}\,,
\end{equation}
where $
\left\langle\, \ldots \,\right\rangle $ stands for the nearest neighbors of $i$-th particle and $N_{i}$ is the number of such neighbors. Thus the vector $\m{u}_{i}$ is just the weighted sum of the random vectors. The number of summands is also random and describes the dynamic coupling between the density and the velocity of the neighbors. Let  one-particle distribution function is $f_{\m{v}}$. Since the order parameter $\m{u}$ is the collective variable as has been said above it changes more slowly than the one-particle function. Thus one can consider $\m{u}$ as the parameter for distribution function $f_{\m{v}}$. Using the standard procedure analogous to the mean-field approach it is possible to get the self-consistent equation for the order parameter. Indeed, assuming that all the correlation between the particles are incorporated into $\m{u}$ we can find the relation between the characteristic functions for $\m{v}$ and $\m{u}$:
\begin{equation}\label{ourmod2}
G_{\m{u}}(\m{k})= \left\langle\, e^{i\m{k}\,\m{u}} \,\right\rangle =\sum _{n=0}^{\infty}p_{n}\,G_{\m{v}}\left(\,\f{\m{k}}{n}\,\right)^{n}\,.
\end{equation}
Here $p_n$ is the probability of $n$ neighbors to appear within the interaction range $R$ and
\begin{equation}\label{charfunct_v}
G_{\m{v}} =
\left\langle\, e^{i\m{k}\,\m{v}} \,\right\rangle
\end{equation}
is the characteristic function of the velocity distribution of
a particle $f_{\m{v}}$. It depends on the average value of the
local order parameter \eqref{localorderparameternorm}. In order
to get the specific results for the order parameter one need to
specify the form of the density distribution for the number of
neighbors and the distribution function $f_{\m{v}}$ for the
velocity. The latter depends on the type of noise introduced
into the motion. The original variant of the CVA
\cite{spp_cva_prl1995,spp_cvv/physa/1999} used so called scalar
noise when the direction of a particle is updated with the
random increment of the angle. In recent work
\cite{spp_gregchate_prl2004} another type of noise, so called
vectorial noise, was proposed. We will consider these type of
noise below.

To simplify the consideration we assume that the probability distribution $p_n$ does not depend on the distribution for the velocity. The simplest choice for the distribution of the number of particles in nearest surrounding is the Poisson distribution:
\begin{equation}\label{scalrnoise1}
p_{n}=\frac{\lambda ^{n}}{n!}\,e^{-\lambda }\,,
\end{equation}
where  $\lambda $  is mean number (density) of particles in the nearest surrounding. In order to get the analytic result in such a case it is expedient to use the non normalized order parameter instead of \eqref{localorderparameternorm}:
\begin{equation}\label{localorderparameter}
   \tilde{\m{u}}_{i}=\sum _{
\left\langle\, i,j \,\right\rangle
}\m{v}_{j}\,.
\end{equation}
Then we get simple expression for the characteristic function $G_{\tilde{\m{u}}}$ :
\begin{equation}\label{poisson_charfunc}
G_{\tilde{\m{u}}}(\m{k})=e^{-\lambda +\lambda\, G_{\m{v}}(\m{k})}\,.
\end{equation}
The standard relation between the moments of the distribution and the derivatives of the characteristic function at $\m{k}$ leads to:
\begin{equation}\label{poisson_uv}
\left\langle\, \tilde{\m{u}} \,\right\rangle =
\lambda \langle \m{v}\rangle \Leftrightarrow   \langle \m{u}\rangle = \langle \m{v}\rangle\,.
\end{equation}
Note that such a simple relation between the order parameter and the particle velocity is due to specific form of the Poisson distribution. It allows to justify the expression for the relaxation rate Eq.~\eqref{A_model} in low density limit. The velocity distribution function depends on the specific way of introducing the source of noise. Below we consider two types of noise which are widely used in simulations.

\subsection{Scalar noise}\label{subsec_scalarnoise}
\begin{figure}
  \includegraphics[scale=0.3]{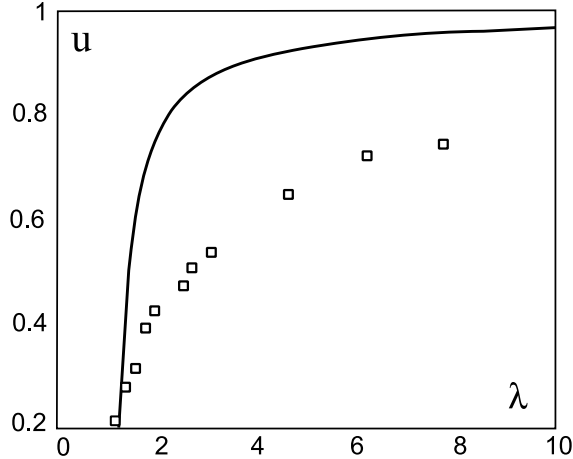}\\
  \caption{Squares are experimental points, solid line is the solution Eq.~\eqref{u_solut}.}\label{scfig2}
\end{figure}

Here we consider the case of so called scalar noise. This is the most obvious way to introduce the stochastic source into the dynamic of a particle. At every step the random increment of the angle is added. Using the results obtained above for the velocity distribution function, we choose it in accordance with Eq.~\eqref{fpstac}

\begin{equation}\label{scalrnoise4}
f_{\m{v}}=C\,e^{\lambda  \f{\,u}{D} \cos{\alpha}}\,,\quad  C=\frac{1}{2\pi I_{0}\left(\lambda  \f{u}{D}\right)}\,.
\end{equation}
\begin{equation}\label{u_solut}
u=\frac{I_{1}\left(\lambda  \f{u}{D}\right)}{I_{0}\left(\lambda  \f{u}{D}\right)}
\end{equation}
It's obvious, that it has trivial solution  $u=0$, which losses its stability depending on the average density $\lambda$ and the diffusion coefficient $D$ (i.e. the intensity of scalar noise). Expanding Eq.~\eqref{u_solut} near the trivial solution we obtain :
\begin{equation}\label{scalrnoise_landauexpansion}
0= \left(\,\f{\lambda}{2D} -1 \,\right)u +\f{\lambda}{16\,D^3}\,u^3 + o(u^3)\,.
\end{equation}
From here the critical density value is as following:
\begin{equation}\label{scalrnoise_lambdacrit}
\lambda _{c}=2\,D
\end{equation}
The comparison of the solution of Eq.~\eqref{u_solut} and  the
results of numerical simulation obtained in
\cite{spp_cva_prl1995,spp_cv/physa/2000} is on
Fig.~\ref{scfig2}.
From Eq.~\eqref{ourmod2} it clear that behavior of the order parameter near the critical value is determined by the analytical properties of the characteristic function $G_{\m{u}}$. Near the critical threshold the dependence of the order parameter has typical for the mean-field approximation square root dependence:
\begin{equation}\label{scalrnoise_sqrt}
u\propto \left(\,\lambda_c - \lambda \,\right)^{1/2}\,.
\end{equation}
The differentiability properties of the characteristic function Eq.~\eqref{ourmod2} and therefore the applicability of the expansion Eq.~\eqref{scalrnoise_landauexpansion} depends on the character of the distribution of the neighbors. If the fluctuations of the density near the threshold value are big this can lead to slow convergence of $p_n$. In such a case one can expect non smooth dependence of $G_{\m{u}}$ on the parameters of the distribution function $f_{\m{v}}$, in particular on the average value of the local order parameter $u$.

In general one need to construct the kinetic equation for the distribution function. Some attempts to derive such equation have been made in a way similar to classic Boltzmann approach \cite{spp_gregorie_kinetic_pre2006} though only binary collisions were taken into account. This approximation is valid only for the system of low density. The applicability of these result to the systems of Vcsek's type is problematic because of the multiparticle character of the ``collision`` process.
\subsection{Vector noise}\label{subsec_vectornoise}
The vectorial noise was introduced in
\cite{spp_gregchate_prl2004} as another realistic model of the
noisy environment. In such a case the random vector $\xi$ is
added, either to the local order parameter $\tilde{\m{u}}$
\cite{spp_gregchate_prl2004} or $\m{u}$
\cite{spp_aldanahuepe_prl2007,spp_aldanamexic_pre2008}. Then
the corresponding direction for the velocity of the particle is
determined:
\begin{equation}\label{vect_noise}
  \theta_{i}(n+1)  = \text{Arg} \left(\, \m{u}_{i}(n) + \boldsymbol \xi_{i} \,\right)
\end{equation}
%
%
In addition, the amplitude of the noise can be chosen so that
$\vert \boldsymbol \xi_{i} \vert = \xi N_{i}$
\cite{spp_gregchate_prl2004}. The results obtained in
\cite{spp_gregchate_prl2004} revealed the difference between
the VM with the scalar noise and raised the intensive
discussion (see
\cite{spp_vicseknagy_pha2007,spp_aldanahuepe_prl2007,spp_aldanamexic_pre2008,spp_gregchaiteanswer2mex_2007}).

Here we derive the one-particle velocity distribution function for the case of vectorial noise and show that it has essentially nonlinear character which leads to the apparent discontinuity in the dependence of the order parameter on the noise intensity.

We assume that the distribution of the direction of the vector $\boldsymbol\xi$ is uniform and independent on the the distribution of the number of neighbors, which is characterized by the corresponding probabilities $p_n$.

\begin{figure}
 \includegraphics[scale=1]{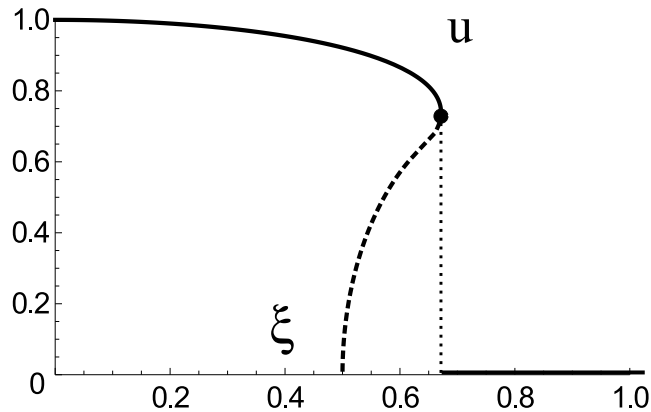}\\
 \caption{Solution of Eq.~\ref{vectnoise_equ}}\label{fig_vectnoisesolut}
\end{figure}

From simple geometrical consideration of vector noise algorithm it is easy to get the relation:
\begin{equation}\label{vectnoise_fn}
\cos \alpha  = \f{1}{\sqrt{1+\f{\sin^2 \varphi}{\left(\,\f{\xi}{u}+\cos{\varphi}\,\right)^2}}}\,,\quad \xi < u\,.
\end{equation}
If $u<\xi$ then the distribution function for the direction is as following:
\begin{equation}\label{falpha}
  f(\alpha) = \f{1}{2\pi}\,\left(\,1+\f{\cos \alpha }{\sqrt{\left(\,\f{\xi}{u}\,\right)^2-\sin^2{\alpha }}}\,\right)
\end{equation}
Thus the self-consistent equation for the order parameter is as following:
\begin{equation}\label{vectnoise_equ}
  u =
\left\langle\, v \,\right\rangle
= F\left(\,\f{\xi}{u}\,\right)
\end{equation}
where
\[
F\left(\,\f{\xi}{u}\,\right) = \left\langle\, \cos \alpha  \,\right\rangle =
  \begin{cases}
    \f{1}{2\pi}\,\intl_{-\pi}^{+\pi} \cos\alpha\,\, \,\left(\,1+\f{1}{\sqrt{\left(\,\f{\xi}{u}\,\right)^2-\sin^2{\alpha }}}\,\right)\,d\alpha \,.
 & \text{if}\,\, u/\xi <1, \\
    \f{1}{2\pi}\intl_{-\pi}^{+\pi}\f{d\varphi}{\sqrt{1+\f{\sin^2 \varphi}{\left(\,\f{u}{\xi}+\cos{\varphi}\,\right)^2}}} & \text{if}\,\, u/\xi > 1.
  \end{cases}\]
The solution of Eq.~\eqref{vectnoise_equ} is shown on
Fig.~\ref{fig_vectnoisesolut}. There is the interval of the
noise intensity $\xi_{1}<\xi<\xi_{2}$, where two nontrivial
solutions for the order parameter exist with the hysteretic (or
subcritical) jump. It is clear that the branch where $du/d\xi >
0 $ represents the unstable state in analogy with the situation
typical for the first order phase transitions. For the model
considered $\xi_1 = 0.5$ and $\xi_2 \approx 0.67$. The
situation here is analogous with that for the Kuramoto model
\cite{spp_kuramotomodel_rpmreview2005}. In the latter case the
type of bifurcation of the partially synchronized phase depends
on the properties of the frequency distribution function
$g(\omega)$ of the oscillators, namely the sign of $g''(0)$
(see \cite{spp_kuramotomodel_rpmreview2005}). Thus we state
that the difference of the one-particle distribution function
in case of scalar and vector noise is the source of the change
the type of the bifurcation from supercritical for scalar noise
to the subcritical for vector noise in the Vicsek model. This
can explain the difference in the results of works
\cite{spp_cva_prl1995} and \cite{spp_gregchate_prl2004}.

%
\section{Conclusion}
In this paper we consider the continuum time limit for the
Vicsek's algorithm \cite{spp_cva_prl1995} of the dynamics of
the self propelling particles. It is shown that there is the
time interval where the kinetic regime of the relaxation of the
particle velocity to the local value of the average velocity of
the neighbors takes place. The relaxation rate depends on the
density linearly at least for not too big number of neighbors.
The cases of vectorial and the scalar noises are considered.
Within the proposed mean field model it is shown that the for
the case of vectorial noise \cite{spp_gregchate_prl2004} the
subcritical bifurcation of the stationary solution takes place.
This is in contrast with the case of scalar noise originally
considered in \cite{spp_cva_prl1995}, where the supercritical
transition occurs.

%
%
%
%

\begin{thebibliography}{21}
\expandafter\ifx\csname natexlab\endcsname\relax\def\natexlab#1{#1}\fi
\expandafter\ifx\csname bibnamefont\endcsname\relax
  \def\bibnamefont#1{#1}\fi
\expandafter\ifx\csname bibfnamefont\endcsname\relax
  \def\bibfnamefont#1{#1}\fi
\expandafter\ifx\csname citenamefont\endcsname\relax
  \def\citenamefont#1{#1}\fi
\expandafter\ifx\csname url\endcsname\relax
  \def\url#1{\texttt{#1}}\fi
\expandafter\ifx\csname urlprefix\endcsname\relax\def\urlprefix{URL }\fi
\providecommand{\bibinfo}[2]{#2}
\providecommand{\eprint}[2][]{\url{#2}}

\bibitem[{\citenamefont{Haken}(1983)}]{book_haken_synergy}
\bibinfo{author}{\bibfnamefont{H.}~\bibnamefont{Haken}},
  \emph{\bibinfo{title}{Synergetics: An Introduction}}
  (\bibinfo{publisher}{Springer Verlag}, \bibinfo{address}{Berlin},
  \bibinfo{year}{1983}).

\bibitem[{\citenamefont{Prigogine and Nicolis}(1977)}]{book_prigoginenicolis}
\bibinfo{author}{\bibfnamefont{I.}~\bibnamefont{Prigogine}} \bibnamefont{and}
  \bibinfo{author}{\bibfnamefont{G.}~\bibnamefont{Nicolis}},
  \emph{\bibinfo{title}{Self-Organization in Non-Equilibrium Systems: From
  Dissipative Structures to Order Through Fluctuations}}
  (\bibinfo{publisher}{J. Wiley \& Sons}, \bibinfo{address}{New York},
  \bibinfo{year}{1977}).

\bibitem[{\citenamefont{Erdmann et~al.}(2005)\citenamefont{Erdmann, Ebeling,
  and Mikhailov}}]{spp_eberdmikh_pre2005}
\bibinfo{author}{\bibfnamefont{U.}~\bibnamefont{Erdmann}},
  \bibinfo{author}{\bibfnamefont{W.}~\bibnamefont{Ebeling}}, \bibnamefont{and}
  \bibinfo{author}{\bibfnamefont{A.}~\bibnamefont{Mikhailov}},
  \bibinfo{journal}{Phys. Rev. E} \textbf{\bibinfo{volume}{71}},
  \bibinfo{pages}{051904} (\bibinfo{year}{2005}).

\bibitem[{\citenamefont{D'Orsogna et~al.}(2006)\citenamefont{D'Orsogna, Chuang,
  Bertozzi, and Chayes}}]{spp_bertoz_prl2006}
\bibinfo{author}{\bibfnamefont{M.~R.} \bibnamefont{D'Orsogna}},
  \bibinfo{author}{\bibfnamefont{Y.~L.} \bibnamefont{Chuang}},
  \bibinfo{author}{\bibfnamefont{A.~L.} \bibnamefont{Bertozzi}},
  \bibnamefont{and} \bibinfo{author}{\bibfnamefont{L.~S.}
  \bibnamefont{Chayes}}, \bibinfo{journal}{Phys. Rev. Lett.}
  \textbf{\bibinfo{volume}{96}}, \bibinfo{pages}{104302}
  (\bibinfo{year}{2006}).

\bibitem[{\citenamefont{Chen and
    Leung}(2006)}]{spp_china_pre2006}
    \bibinfo{author}{\bibfnamefont{H.-Y.} \bibnamefont{Chen}}
    \bibnamefont{and}
  \bibinfo{author}{\bibfnamefont{K.-t.} \bibnamefont{Leung}},
  \bibinfo{journal}{Phys. Rev. E} \textbf{\bibinfo{volume}{73}},
  \bibinfo{pages}{056107} (\bibinfo{year}{2006}).

\bibitem[{\citenamefont{Viscek et~al.}(1995)\citenamefont{Viscek, Czir{\"o}k,
  Ben-Jacob, Cohen, and Shochet}}]{spp_cva_prl1995}
\bibinfo{author}{\bibfnamefont{T.}~\bibnamefont{Viscek}},
  \bibinfo{author}{\bibfnamefont{A.}~\bibnamefont{Czir{\"o}k}},
  \bibinfo{author}{\bibfnamefont{E.}~\bibnamefont{Ben-Jacob}},
  \bibinfo{author}{\bibfnamefont{I.}~\bibnamefont{Cohen}}, \bibnamefont{and}
  \bibinfo{author}{\bibfnamefont{O.}~\bibnamefont{Shochet}},
  \bibinfo{journal}{Phys. Rev. Lett.} \textbf{\bibinfo{volume}{75}},
  \bibinfo{pages}{1226} (\bibinfo{year}{1995}).

\bibitem[{\citenamefont{Kulinskii et~al.}(2005)\citenamefont{Kulinskii,
  Ratushnaya, Zvelindovsky, and Bedeaux}}]{spp_us/eurphyslet/2005}
\bibinfo{author}{\bibfnamefont{V.}~\bibnamefont{Kulinskii}},
  \bibinfo{author}{\bibfnamefont{V.}~\bibnamefont{Ratushnaya}},
  \bibinfo{author}{\bibfnamefont{A.}~\bibnamefont{Zvelindovsky}},
  \bibnamefont{and} \bibinfo{author}{\bibfnamefont{D.}~\bibnamefont{Bedeaux}},
  \bibinfo{journal}{Europhys. Lett.} \textbf{\bibinfo{volume}{71}},
  \bibinfo{pages}{207} (\bibinfo{year}{2005}).

\bibitem[{\citenamefont{Ratushnaya et~al.}(2006)\citenamefont{Ratushnaya,
  Kulinskii, Zvelindovsky, and Bedeaux}}]{spp_usphysicastacsolut/physica/2006}
\bibinfo{author}{\bibfnamefont{V.}~\bibnamefont{Ratushnaya}},
  \bibinfo{author}{\bibfnamefont{V.}~\bibnamefont{Kulinskii}},
  \bibinfo{author}{\bibfnamefont{A.}~\bibnamefont{Zvelindovsky}},
  \bibnamefont{and} \bibinfo{author}{\bibfnamefont{D.}~\bibnamefont{Bedeaux}},
  \bibinfo{journal}{Physica A: Statistical Mechanics and its Applications}
  \textbf{\bibinfo{volume}{366}}, \bibinfo{pages}{107} (\bibinfo{year}{2006}),
  ISSN \bibinfo{issn}{03784371},
  \urlprefix\url{http://dx.doi.org/10.1016/j.physa.2005.11.002}.

\bibitem[{\citenamefont{Sakaguchi et~al.}(1987)\citenamefont{Sakaguchi,
  Shinomoto, and Kuramoto}}]{spp_kuramotosakaguchi/japan/1987}
\bibinfo{author}{\bibfnamefont{H.}~\bibnamefont{Sakaguchi}},
  \bibinfo{author}{\bibfnamefont{S.}~\bibnamefont{Shinomoto}},
  \bibnamefont{and} \bibinfo{author}{\bibfnamefont{Y.}~\bibnamefont{Kuramoto}},
  \bibinfo{journal}{Prog. Theor. Phys.} \textbf{\bibinfo{volume}{77}},
  \bibinfo{pages}{1005} (\bibinfo{year}{1987}).

\bibitem[{\citenamefont{Acebron et~al.}(2005)\citenamefont{Acebron, Bonilla,
  Vicente, Ritort, and Spigleri}}]{spp_kuramotomodel_rpmreview2005}
\bibinfo{author}{\bibfnamefont{J.~A.} \bibnamefont{Acebron}},
  \bibinfo{author}{\bibfnamefont{L.~L.} \bibnamefont{Bonilla}},
  \bibinfo{author}{\bibfnamefont{C.~J.~P.} \bibnamefont{Vicente}},
  \bibinfo{author}{\bibfnamefont{F.}~\bibnamefont{Ritort}}, \bibnamefont{and}
  \bibinfo{author}{\bibfnamefont{R.}~\bibnamefont{Spigleri}},
  \bibinfo{journal}{Rev. Mod. Phys.} \textbf{\bibinfo{volume}{77}},
  \bibinfo{pages}{137} (\bibinfo{year}{2005}).

\bibitem[{\citenamefont{Nagy et~al.}(2007)\citenamefont{Nagy, Daruka, and
  Viscek}}]{spp_vicseknagy_pha2007}
\bibinfo{author}{\bibfnamefont{M.}~\bibnamefont{Nagy}},
  \bibinfo{author}{\bibfnamefont{I.}~\bibnamefont{Daruka}}, \bibnamefont{and}
  \bibinfo{author}{\bibfnamefont{T.}~\bibnamefont{Viscek}},
  \bibinfo{journal}{Physica A} \textbf{\bibinfo{volume}{377}},
  \bibinfo{pages}{445} (\bibinfo{year}{2007}).

\bibitem[{\citenamefont{Czir{\"o}k et~al.}(1996)\citenamefont{Czir{\"o}k,
  Ben-Jacob, Cohen, and Viscek}}]{spp_cva_pre1996}
\bibinfo{author}{\bibfnamefont{A.}~\bibnamefont{Czir{\"o}k}},
  \bibinfo{author}{\bibfnamefont{E.}~\bibnamefont{Ben-Jacob}},
  \bibinfo{author}{\bibfnamefont{I.}~\bibnamefont{Cohen}}, \bibnamefont{and}
  \bibinfo{author}{\bibfnamefont{T.}~\bibnamefont{Viscek}},
  \bibinfo{journal}{Phys. Rev. E} \textbf{\bibinfo{volume}{54}},
  \bibinfo{pages}{1791} (\bibinfo{year}{1996}).

\bibitem[{\citenamefont{Van~Kampen}(2001)}]{book_vankampen}
\bibinfo{author}{\bibfnamefont{N.~G.} \bibnamefont{Van~Kampen}},
  \emph{\bibinfo{title}{Stochastic Processes in Physics and Chemistry
  (North-Holland Personal Library)}} (\bibinfo{publisher}{{North Holland}},
  \bibinfo{year}{2001}), ISBN \bibinfo{isbn}{0444893490}.

\bibitem[{\citenamefont{Csahok and
    Viscek}(1995)}]{spp_cvalattic_pre1995}
    \bibinfo{author}{\bibfnamefont{Z.}~\bibnamefont{Csahok}}
    \bibnamefont{and}
  \bibinfo{author}{\bibfnamefont{T.}~\bibnamefont{Viscek}},
  \bibinfo{journal}{Phys. Rev. E} \textbf{\bibinfo{volume}{52}},
  \bibinfo{pages}{5297} (\bibinfo{year}{1995}).

\bibitem[{\citenamefont{Czir{\"o}k et~al.}(1999)\citenamefont{Czir{\"o}k,
  Viscek, and T.Viscek}}]{spp_cvv/physa/1999}
\bibinfo{author}{\bibfnamefont{A.}~\bibnamefont{Czir{\"o}k}},
  \bibinfo{author}{\bibfnamefont{M.}~\bibnamefont{Viscek}}, \bibnamefont{and}
  \bibinfo{author}{\bibnamefont{T.Viscek}}, \bibinfo{journal}{Physica A}
  \textbf{\bibinfo{volume}{264}}, \bibinfo{pages}{299} (\bibinfo{year}{1999}).

\bibitem[{\citenamefont{Gr\'{e}goire and
  Chat\'{e}}(2004)}]{spp_gregchate_prl2004}
\bibinfo{author}{\bibfnamefont{G.}~\bibnamefont{Gr\'{e}goire}}
  \bibnamefont{and}
  \bibinfo{author}{\bibfnamefont{H.}~\bibnamefont{Chat\'{e}}},
  \bibinfo{journal}{Phys. Rev. Lett.} \textbf{\bibinfo{volume}{92}},
  \bibinfo{pages}{025702} (\bibinfo{year}{2004}).

\bibitem[{\citenamefont{Czir{\"o}k and T.Viscek}(2000)}]{spp_cv/physa/2000}
\bibinfo{author}{\bibfnamefont{A.}~\bibnamefont{Czir{\"o}k}} \bibnamefont{and}
  \bibinfo{author}{\bibnamefont{T.Viscek}}, \bibinfo{journal}{Physica A}
  \textbf{\bibinfo{volume}{281}}, \bibinfo{pages}{17} (\bibinfo{year}{2000}).

\bibitem[{\citenamefont{Bertin et~al.}(2006)\citenamefont{Bertin, Droz, and
  Gr\'{e}goire}}]{spp_gregorie_kinetic_pre2006}
\bibinfo{author}{\bibfnamefont{E.}~\bibnamefont{Bertin}},
  \bibinfo{author}{\bibfnamefont{M.}~\bibnamefont{Droz}}, \bibnamefont{and}
  \bibinfo{author}{\bibfnamefont{G.}~\bibnamefont{Gr\'{e}goire}},
  \bibinfo{journal}{Physical review. E, Statistical, nonlinear, and soft matter
  physics} \textbf{\bibinfo{volume}{74}} (\bibinfo{year}{2006}), ISSN
  \bibinfo{issn}{1539-3755},
  \urlprefix\url{http://view.ncbi.nlm.nih.gov/pubmed/17025488}.

\bibitem[{\citenamefont{Aldana et~al.}(2007)\citenamefont{Aldana, Dossetti,
  Huepe, Kenkre, and Larralde}}]{spp_aldanahuepe_prl2007}
\bibinfo{author}{\bibfnamefont{M.}~\bibnamefont{Aldana}},
  \bibinfo{author}{\bibfnamefont{V.}~\bibnamefont{Dossetti}},
  \bibinfo{author}{\bibfnamefont{C.}~\bibnamefont{Huepe}},
  \bibinfo{author}{\bibfnamefont{V.~M.} \bibnamefont{Kenkre}},
  \bibnamefont{and} \bibinfo{author}{\bibfnamefont{H.}~\bibnamefont{Larralde}},
  \bibinfo{journal}{Physical Review Letters} \textbf{\bibinfo{volume}{98}},
  \bibinfo{eid}{095702} (pages~\bibinfo{numpages}{4}) (\bibinfo{year}{2007}),
  \urlprefix\url{http://link.aps.org/abstract_prlv98/e095702}.

\bibitem[{\citenamefont{Pimentel et~al.}(2008)\citenamefont{Pimentel, Aldana,
  Huepe, and Larralde}}]{spp_aldanamexic_pre2008}
\bibinfo{author}{\bibfnamefont{J.~A.} \bibnamefont{Pimentel}},
  \bibinfo{author}{\bibfnamefont{M.}~\bibnamefont{Aldana}},
  \bibinfo{author}{\bibfnamefont{C.}~\bibnamefont{Huepe}}, \bibnamefont{and}
  \bibinfo{author}{\bibfnamefont{H.}~\bibnamefont{Larralde}},
  \bibinfo{journal}{Physical Review E (Statistical, Nonlinear, and Soft Matter
  Physics)} \textbf{\bibinfo{volume}{77}} (\bibinfo{year}{2008}),
  \urlprefix\url{http://dx.doi.org/10.1103/PhysRevE.77.061138}.

\bibitem[{\citenamefont{Chat\'{e} et~al.}(2007)\citenamefont{Chat\'{e},
  Ginelli, and Gr\'{e}goire}}]{spp_gregchaiteanswer2mex_2007}
\bibinfo{author}{\bibfnamefont{H.}~\bibnamefont{Chat\'{e}}},
  \bibinfo{author}{\bibfnamefont{F.}~\bibnamefont{Ginelli}}, \bibnamefont{and}
  \bibinfo{author}{\bibfnamefont{G.}~\bibnamefont{Gr\'{e}goire}},
  \bibinfo{journal}{Physical Review Letters} \textbf{\bibinfo{volume}{99}},
  \bibinfo{eid}{229601} (pages~\bibinfo{numpages}{1}) (\bibinfo{year}{2007}),
  \urlprefix\url{http://link.aps.org/abstract_prlv99/e229601}.

\end{thebibliography}
\end{document}